\documentclass[runningheads]{llncs}

\usepackage{listings}
\usepackage{eccv}

\usepackage{eccvabbrv}

\usepackage{graphicx}
\usepackage{booktabs}
\usepackage{wrapfig,lipsum,booktabs}
\usepackage[accsupp]{axessibility}  
\usepackage{xcolor}
\usepackage[misc]{ifsym}
\definecolor{codegreen}{rgb}{0,0.6,0}
\definecolor{codegray}{rgb}{0.5,0.5,0.5}
\definecolor{codepurple}{rgb}{0.58,0,0.82}
\definecolor{backcolour}{rgb}{0.95,0.95,0.92}

\lstdefinestyle{mystyle}{
    backgroundcolor=\color{backcolour},   
    commentstyle=\color{codegreen},
    keywordstyle=\color{magenta},
    numberstyle=\tiny\color{codegray},
    stringstyle=\color{codepurple},
    basicstyle=\ttfamily\footnotesize,
    breakatwhitespace=false,         
    breaklines=true,                 
    captionpos=b,                    
    keepspaces=true,                 
    numbers=left,                    
    numbersep=5pt,                  
    showspaces=false,                
    showstringspaces=false,
    showtabs=false,                  
    tabsize=2
}

\usepackage{hyperref}

\usepackage{orcidlink}

\begin{document}

\title{Differentiable Convex Polyhedra Optimization from Multi-view Images } 

\titlerunning{DiffConvex}

\author{Daxuan Ren\inst{1} \and
Haiyi Mei, \inst{2} \and 
Hezi Shi\inst{1} \and \\
Jianmin Zheng\textsuperscript \Letter \inst{1} \and
Jianfei Cai\inst{1, 3} \and
Lei Yang \inst{2}}
\authorrunning{D. Ren et al.}

\institute{College of Computing \& Data Science,\;Nanyang Technological University,\;Singapore \and
Sensetime Research \and Department of Data Science \& AI, Monash University
Monash University\\
\email{\{daxuan001, hezi001\}@e.ntu.edu.sg,
\{meihaiyi, yanglei\}@sensetime.com, \\
\ asjmzheng@ntu.edu.sg, jianfei.cai@monash.edu}}

\maketitle

\begin{center}
    \centering
    \captionsetup{type=figure}
    \includegraphics[width=0.99\textwidth]{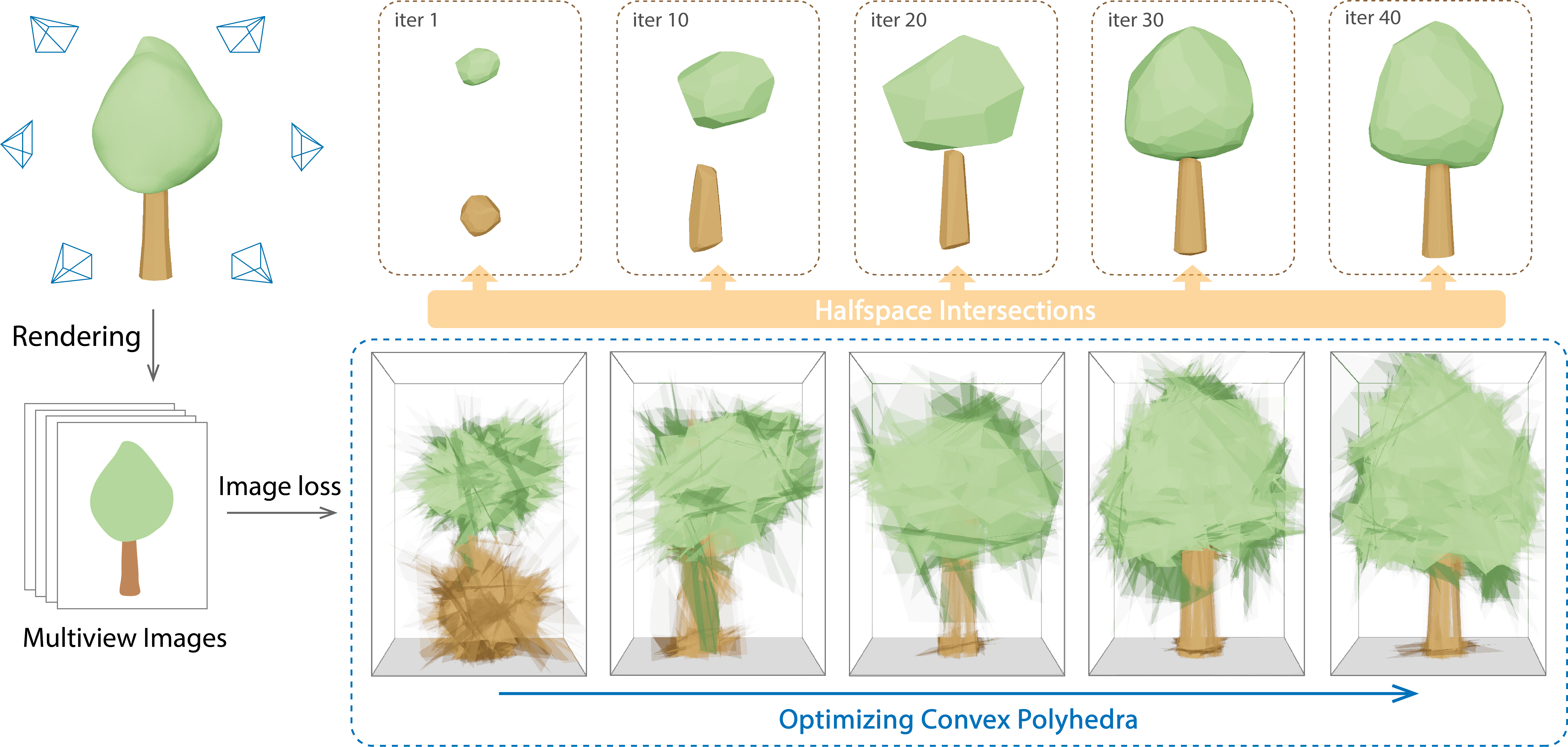}
    \captionof{figure}{A novel method optimizes differentiable convex polyhedra w.r.t image losses, bridging the gap between compact shape representation and easily obtained image supervision.}
\end{center}%

\begin{abstract}
This paper presents a novel approach for the differentiable rendering of convex polyhedra, addressing the limitations of recent methods that rely on implicit field supervision. Our technique introduces a strategy that combines non-differentiable computation of hyperplane intersection through duality transform with differentiable optimization for vertex positioning with three-plane intersection, enabling gradient-based optimization without the need for 3D implicit fields. This allows for efficient shape representation across a range of applications, from shape parsing to compact mesh reconstruction. This work not only overcomes the challenges of previous approaches but also sets a new standard for representing shapes with convex polyhedra. 
\end{abstract}

\section{Introduction}

The quest for shape representation in computer graphics and computer vision has led to significant advancements, exploring various paradigms such as point clouds, voxels, meshes, implicit fields, and geometric primitives. Among these, the technique of using a combination of simple geometric primitives to represent more complex shapes is a promising way of capturing the geometric essence of objects ~\cite{tulsiani2017learning, paschalidou2019superquadrics, chen2020bsp, deng2020cvxnet, ren2021csg, ren2022extrudenet, yu2021capri, yu2023dualcsg}. Particularly, representing shapes via a set of convex polyhedra has shown potential in many downstream applications ~\cite{lien2004approximate, lien2007approximate},  including parsing shapes to semantically meaningful pieces, efficient physics-based simulations, collision detection, and compact mesh representations for efficient rendering and storage. Despite its potential, a fundamental challenge remains: how to enable the differentiable construction and optimization of convex polyhedra. This capability is crucial for leveraging gradient-based optimization techniques, thus enhancing the adaptability and precision of shape representations within learning-based frameworks.

Recent advances in shape representation with primitives have predominantly relied on supervision from implicit fields \cite{chen2020bsp, deng2020cvxnet,kania2020ucsg, ren2021csg, ren2022extrudenet, yu2021capri, yu2023dualcsg, li2023secad}, such as Signed Distance Functions (SDF) or occupancy fields. While effective, these methods necessitate watertight meshes for training, which can be difficult to acquire. Apart from this, preprocessing the dataset into a set of discrete sample points with ground truth implicit values imposes a huge burden on both computation resources and storage. Additionally, during training, all of these methods define shape compositing operations such as intersecting planes and uniting convex polyhedra in approximated ways, like sigmoid for converting SDF into occupancy, and softmin and softmax for intersection and union, leading to inaccurate overall SDF or occupancy fields or biased gradient estimate. Also, since the gradient of each sample point w.r.t all the primitives needs to be cached during training, the consumption of GPU memory also imposes a barrier for scaling up the models.

This paper presents an approach that provides a new direction for eliminating the dependency on implicit fields. It is grounded on purely explicit surface models and leverages images rendered in a differentiable manner for supervision. This strategy offers a more flexible solution for gradient-based optimization of shape representation through a collection of convex polyhedra.

Specifically, our method combines highly efficient non-differentiable computations with the adaptability of differentiable operations. Utilizing the duality transform concept~\cite{preparata2012computational, de2000computational}, we precisely identify the hyperplanes that form each vertex of a convex polyhedron—an inherently non-differentiable task. With the hyperplane intersections mapped to each vertex of the polyhedron, we then employ a differentiable linear system solver to calculate the vertex positions by solving for the intersection of three planes. This approach ensures the seamless backpropagation of gradients from image-based losses to the plane parameters defining the convex polyhedron, making the whole optimization process differentiable. Since our method only requires images as training data, it removes the requirements for ground truth meshes, unlocking training data with orders of magnitude larger size. 

In our approach, each vertex location is directly solved explicitly as three-plane intersection, which makes the reconstructed mesh much more accurate in representing the shape compared to previous methods that leverage implicit fields, and offers more accurate gradients around shape boundaries as well. 

Consequently, our method paves the way for more efficient and scalable shape representation learning, making it particularly advantageous for applications where high fidelity is required.

We showcase the broad applicability and superior effectiveness of our method through its successful deployment in a diverse array of applications. Our technique excels in shape reconstruction, where it achieves high-fidelity representations of complex geometries, especially on parts with detailed geometry. In textured multiview reconstruction, it seamlessly integrates diverse visual perspectives into coherent, detailed models, underscoring its ability to handle varied data inputs. Additionally, our approach demonstrates its strength in shape parsing by accurately decomposing complex shapes into semantically meaningful components. All these applications illustrate not only the versatility of our method but also its capacity to deliver enhanced performance and efficiency. 

In summary, this paper has the following contributions:
\begin{itemize}
    \item We introduce a novel method for making the construction of convex polyhedra differentiable, enabling gradient-based optimization without reliance on implicit fields.
    \item Our approach combines non-differentiable computation via duality transform with differentiable optimization, allowing for effective and accurate shape representation.
    \item  We demonstrate the versatility and efficiency of our method across a range of applications, from shape reconstruction, textured multiview reconstruction, to shape parsing.
\end{itemize}

\section{Related Work}

This section briefly discusses related topics, focusing on shape representations, especially with geometric primitives, and differentiable rendering.

\subsection{Shape Representations}
\noindent \textbf{Voxels} directly expand pixels from 2D images into 3D space, representing shapes with a regular grid of cubic cells (voxels). Each cell can store information such as material properties, occupancy, or color. This representation is well-suited for volumetric data, offering straightforward manipulation and visualization. However, it is often limited by high memory consumption \cite{maturana2015voxnet, noh2015learning,qi2016volumetric, wang2017cnn}.

\noindent \textbf{Point clouds} represent a shape as a collection of discrete points in space, capturing the surface geometry without explicit connectivity information. This representation is widely used in 3D scanning and reconstruction, where raw data typically form a point cloud. Despite its popularity, the lack of surface connectivity requires additional processing for applications that demand explicit surface models. Recently, many methods have been proposed for various tasks based on point clouds, ranging from shape understanding  ~\cite{qi2017pointnet, qi2017pointnet++, wang2019dynamic}, shape reconstruction~\cite{xu2022point} to shape generation\cite{nichol2022point}.

\noindent \textbf{Meshes} 
use vertices, edges, and faces in a connected topology to represent shapes. Triangular and quadrilateral meshes are the most common, offering detailed and flexible representations of complex surfaces. Meshes are a cornerstone in modern computer graphics for modeling, rendering, animation, and geometric analysis~\cite{blender,wang2018pixel2mesh,wen2019pixel2mesh++, guo20153d, Nicolet2021Large}.

\noindent \textbf{Implicit fields} (or implicit surfaces) define shapes through scalar fields, with the surfaces represented as level sets of the implicit functions (e.g., the set of points whose signed distance function values are zero). This representation supports complex topologies and smooth surfaces, and is also favored for blending and modeling smooth transitions between shapes. In recent years, leveraging implicit fields has become a hot topic for its capability of representing smooth shapes with complicated topology and easy integration into deep learning frameworks. Implicit fields have been widely used for shape learning \cite{mescheder2019occupancy, park2019deepsdf, hao2020dualsdf} and differentiable rendering in computer vision \cite{mildenhall2020nerf, yariv2021volume, wang2021neus, Yu2022MonoSDF, wu2022objectsdf}.

\noindent \textbf{Hybrid Approaches} represent shapes using both an implicit field and a surface mesh via differentiable iso-surface extraction methods. For instance, Deep Marching Cubes~\cite{liao2018deep} adapts the classic marching cubes algorithm for differentiable use, enabling the extraction of surface geometry from volumetric representations through learnable parameters. Deep Marching Tetrahedra ~\cite{shen2021deep} further extends this approach by introducing differentiability into the process of converting scalar fields to meshes via deformable tetrahedron grids. This facilitates the optimization of complex geometries using gradient descent. FlexiCube~\cite{shen2023flexible} proposes a differentiable iso-surface extraction method based on Dual Marching Cubes, offering efficient and flexible shape optimization.

\noindent \textbf{Primitive-based representations}\label{sec:related_primitive_representation}
 model shapes by combining basic geometric entities such as planes, spheres, cubes, cylinders, quadrics, superqudrics~\cite{tulsiani2017learning, paschalidou2019superquadrics}. While they might not be as flexible or might not capture as much detail as other representations, they offer advantages on shape parsing, compactness, editability, etc. One prominent representation is Constructive Solid Geometry (CSG)~\cite{deng2020cvxnet, chen2020bsp, ren2021csg, yu2021capri, yu2023dualcsg, ren2022extrudenet, li2023secad}, which models complex objects by combining simple solid primitives using Boolean operations (union, intersection, difference). Among the existing methods, CVXNet~\cite{deng2020cvxnet} and BSPNet~\cite{chen2020bsp} are most similar to ours. Both of them represent shapes implicitly as the union of a set of convex polyhedra, which result from intersecting hyperplanes.

\subsection{Differentiable Rendering}\label{sec:related_diff_rendering}
Differentiable rendering bridges the gap between 3D model representation and image-based loss functions, allowing for the optimization of shape parameters directly from image data. Differentiable rendering can be roughly categorized into three categories: rasterization-based, physical-based, and implicit field based. Rasterization based differentiable renderers~\cite{ravi2020pytorch3d, liu2019softras, Laine2020diffrast, munkberg2022extracting} generally offer fast rendering, which use a ``soft'' version of rasterization and shading to allow gradients to propagate, however it generally only support explicit triangle mesh as shape representation. Unlike rasterization based differentiable renderers, physical-based differentiable renderers are built on top of ray tracing techniques and support a border spectrum of light transport phenomena such as reflection, refraction, and transparent objects \cite{li2018differentiable, loubet2019reparameterizing, nimier2019mitsuba, bangaru2020unbiased,  Mitsuba3}, these methods can support shape representation other than triangle mesh, as long as explicit boundaries or a continuous and boundary consistent reparameterization is provided. With the research advancement in the field of implicit fields, the renders for implicit fields \cite{mildenhall2020nerf, wang2021neus, yariv2021volume, jiang2020sdfdiff, Vicini2022sdf, bangaru2022differentiable} also become popular. Typically, these methods use volume rendering to integrate radiance along the ray direction and employ alpha compositing to determine each pixel's color. Additionally, 3DGS ~\cite{kerbl20233d} proposed a hybrid approach that represents scenes with gaussian balls, combining rasterization with volumetric rendering to generate images. These methods have recently gained significant attention due to the flexibility and continuous nature of implicit fields. However, none of these method can be applied directly to in our case, since: \textbf{1)} there is no explicit surface mesh available, \textbf{2)} a continuous and boundary-consistent reparameterization is hard to design, especially with potential self-intersecting convex polyhedra, and \textbf{3)} volume rendering require untractable amount of VRAM when paired with the primitive-based representations discussed in Sec.\ref{sec:related_primitive_representation}. For a 512x512 image with 64 convex polyhedra, each defined by 64 hyperplanes, and each ray sampling 512 points for ray integration, the total number of queries per image becomes impractically high (512x512x64x64x512 = 549,755,813,888). This exceeds the capacity of contemporary GPU memory. These challenges motivated the approach described in this work.

\section{Methods}
\begin{figure}[ht]
    \centering
    \includegraphics[width=0.98\textwidth]{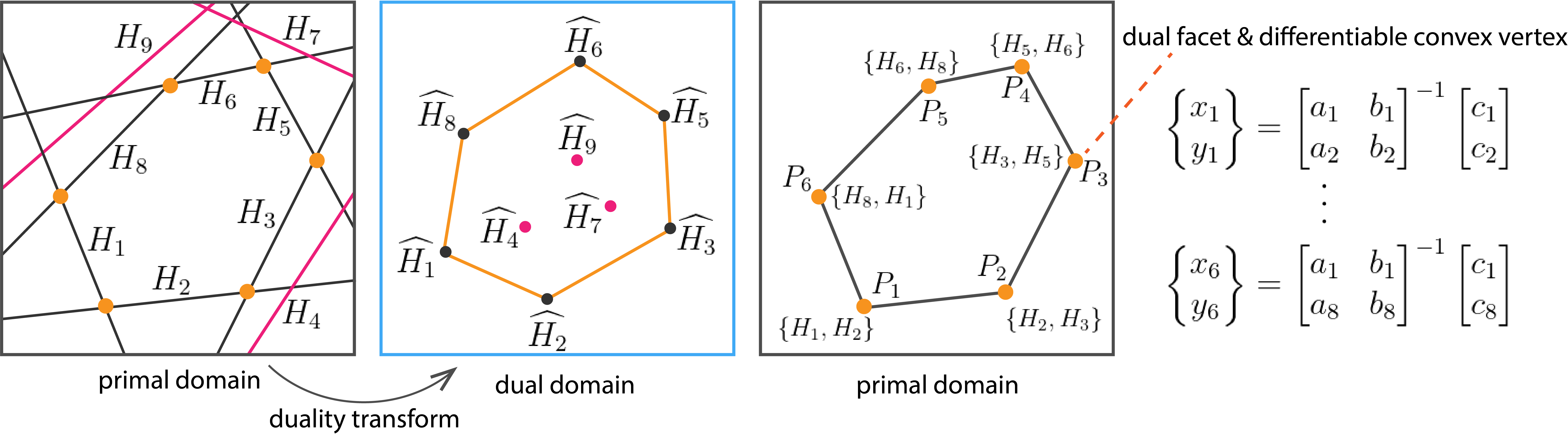}
    \caption{Overview of the proposed method. Given a set of hyperplanes, we use duality transform to map them into the dual space. We then compute the convex hull of the dual vertices. Each facet of the dual convex hull represent a intersection point in the primal domain. Once the plane IDs for each intersection vertex are recorded, we can recompute the vertex location via differentiable linear equation solvers.}
    \label{fig:overview}
\end{figure}

\noindent\textbf{Overview}. 
The essence of our approach is to employ the intersections and unions of a set of halfspaces to depict shapes. To equip these shapes with differentiability, we initially construct convex polyhedra from the intersections of halfspaces using non-differentiable duality transforms and record the IDs of three planes that intersect to form each vertex. Subsequently, we recompute the vertex positions of polyhedra through a differentiable process for finding the intersection of three planes. This enables the vertex positions to be differentiable, allowing the straightforward application of a differentiable renderer  to calculate image loss and facilitating the backpropagation of gradients to the plane parameters. Notably, the union of convex polyhedra is seamlessly managed by the renderer, utilizing mechanisms such as the z-buffer or ray-object intersections, thereby obviating the need for explicit handling. Fig.~\ref{fig:overview} illustrates these processes.

\subsection{Differentiable Rendering of Convex Polyhedra}
Differentiable rendering of convex polyhedra involves the computation of gradients of plane parameters $\mathcal{\theta}$ w.r.t image loss $\mathcal{L}$: $\frac{\partial \mathcal{L}}{\partial \mathcal{\theta}} = \frac{\partial \mathcal{L}}{\partial \mathcal{I}} \cdot \frac{\partial \mathcal{I}}{\partial \mathcal{\theta}}$.

However, directly evaluating $\frac{\partial \mathcal{I}}{\partial \mathcal{\theta}}$ is difficult, see Sec.\ref{sec:related_diff_rendering}. Thus, we choose an indirect approach by constructing the mesh of the convex polyhedron where the vertex position 
is differentiable w.r.t plane parameters. After this, we can use standard differentiable renderer to obtain the 
final gradients:
\begin{equation}
    \frac{\partial \mathcal{L}}{\partial \mathcal{\theta}} = 
    \underbrace{ \frac{\partial \mathcal{L}}{\partial \mathcal{I}} }_\text{Differentiable Image Loss}
    \cdot \underbrace{ \frac{\partial \mathcal{I}}{\partial \mathcal{V}} }_\text{Differentiable Rendering}
    \cdot \underbrace{ \frac{\partial \mathcal{V}} {\partial \mathcal{\theta}} }_\text{Differentiable Vertex Location}
\end{equation}

\subsection{Halfspace Intersections about a Point}
\label{sec:halfspace-intersection}
In Euclidean space $\mathbb{R}^n$, a halfspace is the set of points $x \in \mathbb{R}^n$ satisfying a linear inequality of the form $a^Tx \leq b$, where $a \in \mathbb{R}^n$ is a nonzero vector, $b \in \mathbb{R}$ is a scalar, and $a^Tx$ represents the dot product of $a$ and $x$.
 
A convex set (polyhedron) is a subset of $\mathbb{R}^n$ that, for every pair of its points, the line segment connecting the two points is contained in the set. Formally, a set $C \subseteq \mathbb{R}^n$ is convex if for every $x, y \in C$ and every $\lambda \in [0, 1]$, the point $\lambda x + (1 - \lambda)y$ is also in $C$.

The convex set generated by intersecting a set of halfspaces is defined as follows.
Let $\{H_i\}_{i=1}^k$ be a collection of halfspaces in $\mathbb{R}^n$, where each halfspace $H_i$ is defined by a linear inequality $a_i^Tx \leq b_i$ with $a_i \in \mathbb{R}^n$ and $b_i \in \mathbb{R}$. The intersection of these halfspaces, denoted by $C$, is defined as:
\begin{equation}\label{eq:half_space_intersection}
    C = \bigcap_{i=1}^k H_ = \left\{ x \in \mathbb{R}^n \mid a_i^Tx \leq b_i, \, \forall i = 1, \ldots, k \right\}.
\end{equation}

\noindent\textbf{Duality Transform}.
 The duality transform is a powerful tool in computational geometry that allows us to view the problem from a different perspective. To give a concrete example, a line in 2D Euclidean space has two parameters, i.e., the slope $m$ and the Y-intercept $n$. Therefore, we can draw a point in the dual domain with coordinate $(m,n)$. This mapping between the line in the primal domain and the point in the dual domain (and vice versa) is known as the duality transform. Duality transform can be very useful in many computation geometry applications, such as the construction of convex hull, Delaunay triangulation, Voronoi diagram, and in our case, halfspace intersections. 

\noindent\textbf{Halfspace Intersection with Duality Transform}.
 Given a set of halfspaces defined by linear inequalities $a_i^Tx \leq b_i$ for $i = 1, \ldots, k$, and a feasible point $x_0$ that satisfies all these inequalities, the duality transform involves converting each halfspace into a point in the dual domain and vice-versa. This transformation facilitates the identification of all vertices generated by intersecting the halfspaces. Specifically, the dual of a hyperplane $a_i^Tx = b_i$ 
 is defined by a point with coordinates $(\frac{a_{i_x}}{b_i}, \frac{a_{i_y}}{b_i}, \frac{a_{i_z}}{b_i})$ in the dual domain. By computing the convex hull of these points in the dual domain, we can effectively identify the boundaries of the intersection of halfspaces in the primal domain.  Notably, each facet of the dual convex hull corresponds to a vertex in the primal domain formed by intersecting the hyperplanes corresponding to the vertices of the dual facet~\cite{preparata2012computational}. 
 This approach can significantly simplify the computation by converting a potentially complex halfspace intersection problem into a convex hull problem in the dual domain. Once the three planes that intersect into each convex polyhedron vertex is know, the position of the vertex can be computed differentially via the solution of a system of linear equations. Specifically, if a vertex in the primal space is formed by the intersection of $n$ hyperplanes, then the position of this vertex, denoted by $x$, can be found by solving the system: $ A x = b$ 
where $A$ is a matrix whose rows are the normal vectors of the hyperplanes $a_i^T$ for $i = 1, \ldots, n$, and $b$ is a vector containing the offsets of these hyperplanes, with each entry $b_i$ corresponding to the distance from the origin to the hyperplane. Formally, this system can be written as:

\begin{equation}
\begin{bmatrix}
a_1^T \\
\vdots \\
a_n^T
\end{bmatrix} x = 
\begin{bmatrix}
b_1 \\
\vdots \\
b_n
\end{bmatrix}.
\end{equation}
The solution to this system gives the coordinates of the vertex in the primal space. Since solving the linear system can be done in a differentiable fashion, $\frac{\partial \mathcal{V}} {\partial \mathcal{\theta}}$ can be obtained. 

Having tackled vertex computation, we shift our attention to triangle face generation. As optimizing the connectivity of vertices often falls outside the scope of most differentiable renderers, we opt for directly utilizing the facets of the convex hull constructed in the primal space through non-differentiable methods and tessellate them into triangles. After computing the vertices and triangle faces of a convex polyhedron, the mesh is fed into a differentiable renderer for rendering.

\begin{figure}[t]
    \centering
    \includegraphics[width=0.98\textwidth]{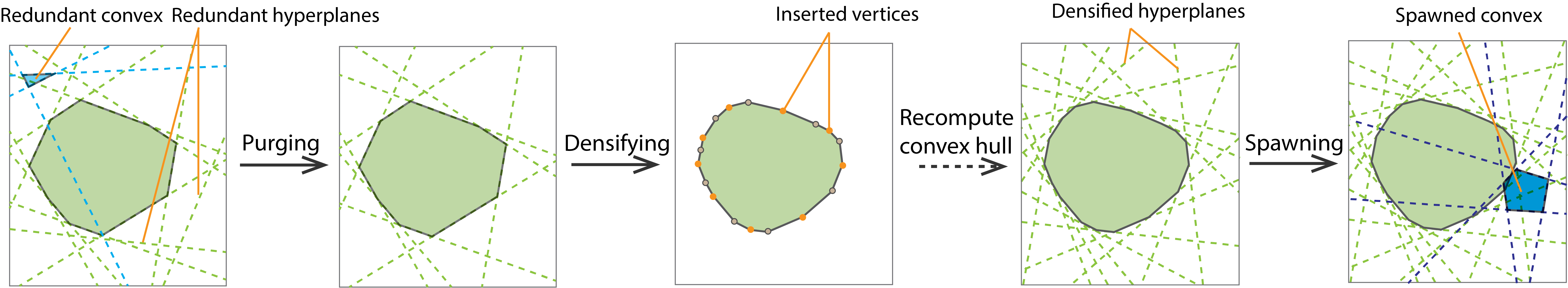}
    \caption{Small convex polyhedra and redundant planes will be removed to speed up the optimization process. To better reconstruct the shape with high curvature, we employ a densification process that constructs the mesh of the convex polyhedron, runs Loop subdivision, and uses the recomputed convex hull equations of the subdivided mesh to serve as the updated plane parameters.}
    \label{fig:convex_ops}
\end{figure}

\subsection{Optimization Strategies}\label{sec:optimization}
Having discussed the construction of an individual convex polyhedron, now we turn to optimizing a set of convex polyhedra to represent a scene with complex geometries. Inspired by 3DGS ~\cite{kerbl20233d}, we adapted several heuristics during optimization.  Fig.\ref{fig:convex_ops} illustrates an overview of operations on convex polyhedra.

\noindent \textbf{Convex Initialization}.
 We commence our scene construction with a predetermined quantity of convex polyhedra and planes. The initialization process assigns a uniform size to all convex polyhedra, denoted by the same $b$ value as outlined in Eq.~\ref{eq:half_space_intersection}. Convexes are randomly initialized throughout the space.
 
\noindent \textbf{Persistent Convex}. 
To facilitate the intersection of hyperplanes through duality, our method begins by securing a feasible starting point. Although identifying a feasible solution such as determining the Chebyshev center of a linear programming problem is relatively straightforward, we circumvent this necessity by ensuring that the origin is always within the convex hull. This is accomplished by introducing additional constraints to Eq.~\ref{eq:half_space_intersection}, mandating that all plane offsets remain positive ($b\in\mathbb{R^+}$). This strategy ensures that the origin acts as a feasible solution. Moreover, to provide spatial flexibility, we incorporate an extra translation parameter for each convex. This adjustment allows for the seamless relocation of the entire convex polyhedron within the three-dimensional space, enhancing the adaptability of our approach.

\noindent \textbf{Convex Purging}. 
The treatment to convex polyhedra introduced above guarantees their persistence by consistently enclosing a viable region, thus preventing from the elimination of small convex polyhedra throughout the optimization process. To improve both the visual quality and computational efficiency, we adopt a convex purging strategy. This strategy entails the removal of convex polyhedra along with their associated parameters once their volumes are below a certain predefined threshold.

\noindent \textbf{Plane Purging}. 
During optimization, if a plane does not contribute to the computation of any vertex, it will not receive any gradient and never get updated. Thus, we further prune inactive hyperplanes to expedite the optimization. This can be done by removing all the hyperplanes that have a dual vertex falling within the dual convex hull. 

\noindent\textbf{Convex Densification}.
 Given the initialization of the convex polyhedron with a relatively small number of planes, our approach initially struggles to accurately reconstruct regions exhibiting high curvature. To address this limitation, we employ a densification strategy that reintroduces the necessary hyperplanes, and provides a balance between model sparsity and the need for detailed shape reconstruction. Specifically, after a predetermined number of optimization iterations, we activate a convex densification process aimed at enhancing the reconstruction of areas with high curvature. This process is executed through a method akin to mesh subdivision: for each convex polyhedron, we calculate the triangle mesh of the convex polyhedron, apply one Loop subdivision iteration, and then recompute the convex hull of the subdivided mesh. The plane equations derived from the new convex hull are then utilized as the parameters for the planes in the subseqent optimization iterations, ensuring a more accurate and comprehensive representation of complex shapes.

\noindent \textbf{Convex Spwaning}. 
Convex purging, the process of discarding small convex polyhedra, potentially risks entrapment in local minima and the loss of details. To counteract this and ensure that a sufficient number of convex polyhedra are employed for the final shape reconstruction, we introduce a convex spawning operation. This procedure involves randomly re-initializing removed convex polyhedra within the space. Through practical applications, we have found that this random spawning approach is effective in maintaining an adequate count of convexes for the final reconstruction. Nonetheless, we acknowledge the possibility of more sophisticated spawning strategies that could enhance this process, suggesting a direction for future research.

\section{Experiments}

\subsection{Shape Reconstruction} \label{sec:shape_reconstruction}
\noindent\textbf{Dataset}
We begin experiments with multiview reconstruction using the ShapeNet dataset~\cite{chang2015shapenet}. Our study focuses on a standard subset of 13 categories. Given that our method does not include a learning component, we evaluate our approach on 100 randomly selected shapes from each category, constrained by computational resource limitations. For each shape, we render 16 images from different viewpoints to act as supervision data.

\begin{figure}[h]
    \centering
    \includegraphics[width=0.98\textwidth]{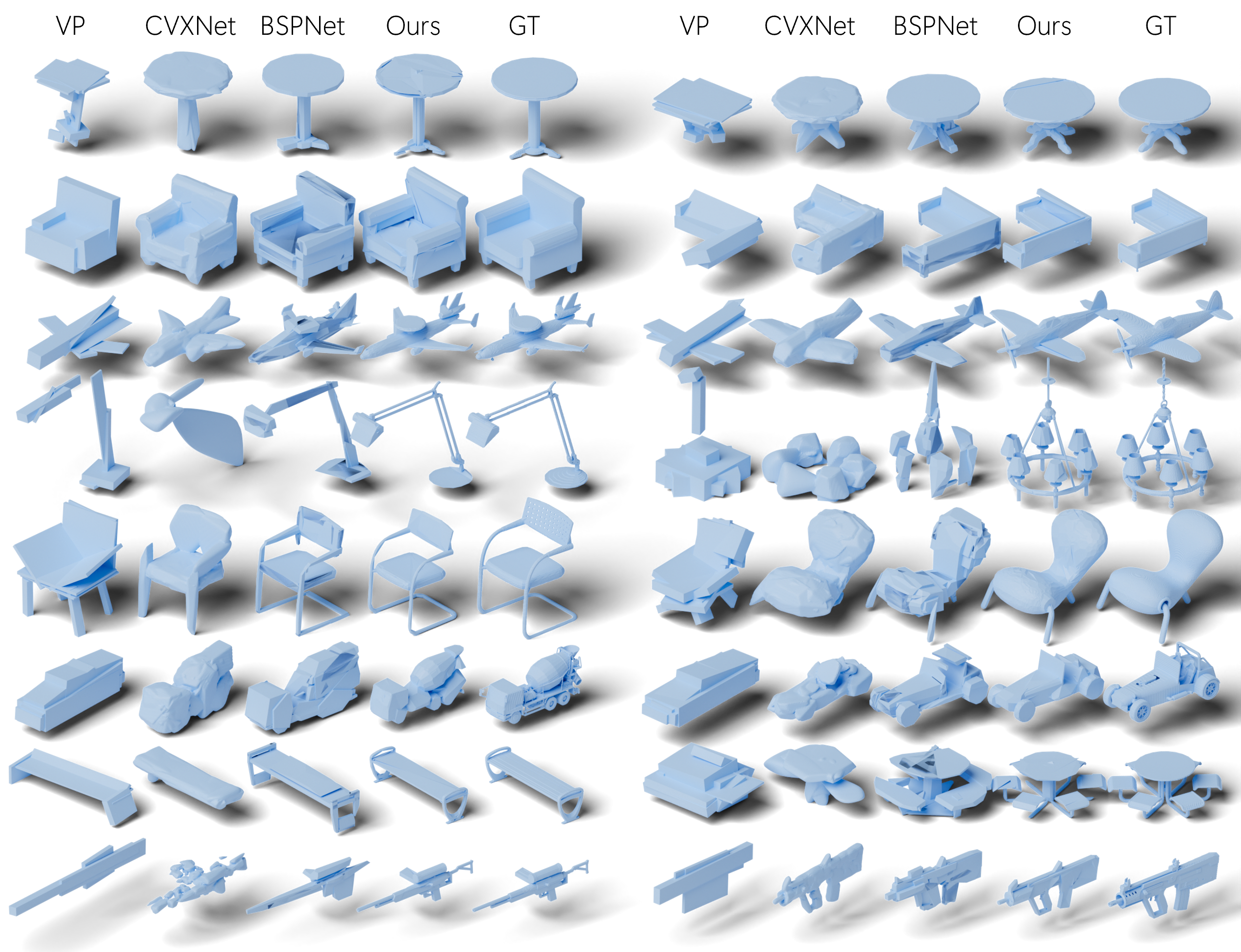}
    \caption{Comparing our method with VP, CVXNet, and BSPNet on the ShapeNet dataset. The visualization results show that our method generates better reconstruction, especially in the thin and detailed aspects of shapes.}
    \label{fig:compare_baselines}
\end{figure}

\noindent\textbf{Implementation Details} \label{sec:implemtation_details}
We implement our method using C++ with QHull for duality related computation. To facility gradient base optimization, we implemented a Python wrapper for it. Once the plane IDs for each convex polyhedron vertex is outputted, we use PyTorch's linear algebra solver to determine the vertex position. We use a total of 32 convex polyhedra for our experiment.   
We randomly initialize the convex polyhedra in space and proceed the optimization steps as detailed in Sec.\ref{sec:optimization}. We optimize a total of 20000 steps, incorporating 10 densification and random spawning steps to refine our model further. We use NVDiffRast~\cite{Laine2020diffrast} as our renderer and build our pipeline based on ~\cite{loubet2019reparameterizing}, due to the code readability and ease of integration. Note that our implementation does not receive any specialized treatments in ~\cite{loubet2019reparameterizing} such as re-parameterization. Instead, we simply reuse their rendering setup such as lighting and shading. Our pipeline is supervised by purely image $L_1$ loss. We use a learning rate of $1e^{-2}$ for convex polyhedron translation and $1e^{-3}$ for hyperplanes.

\noindent\textbf{Comparisons}
In our evaluation, we benchmark our method against other methods focused on shape representation through basic geometric primitives, such as VP\cite{tulsiani2017learning}with 32 cuboids (increased compare to their original paper), CVXNet~\cite{deng2020cvxnet} with 64 convexes (same as their paper), and BSPNet~\cite{chen2020bsp} with 64 convexes (same as their paper). Given that our approach does not have a ``learning'' component and is designed to optimize convex parameters to ``overfit'' to individual shapes, the metric of the baseline method is also ``overfit'' to the dataset, i.e. using training set as test set. Acknowledging our method's lack of visual supervision for internal structures, we employ a postprocessing step from OccNet \cite{mescheder2019occupancy} that sets random cameras, renders multiple depth image and fuses them into a final surface mesh. We use this method to effectively remove any internal structure from both ground truth and output shapes, facilitating fair comparisons.

In Tab.\ref{table:compare_baselines}, we detail our evaluation through quantitative metrics, i.e., $L_1$ Chamfer Distance, $L_2$ Chamfer Distance  (multiplied by 1000 for better viewing), and Normal Consistency. Our method outperforms the baselines on most benchmarks. Qualitative comparisons are shown in Fig.\ref{fig:compare_baselines}. These assessments demonstrate that our method excels in the precise reconstruction of thin and intricate details, for example, the Lamp category in Fig.\ref{fig:compare_baselines}. However, we observe performance discrepancies between $L_1$ and $L_2$ Chamfer Distances, caused by some floating convexes that have not been removed by the optimization. Additionally, for categories with large cavities (e.g., cabinets), our method does not perform as well, likely due to occlusion when relying solely on RGB image supervision.

\begin{table*}[t!]
\caption{Comparison of reconstruction results with baselines, measured by $L_1$ Chamfer Distance, $L_2$ Chamfer Distance, and Normal Consistency. Our method outperforms the baselines and achieves the best overall reconstruction results.}
\begin{center}
\resizebox{0.99\linewidth}{!}{
\addtolength{\tabcolsep}{8pt}   
\begin{tabular}{l |rrrr| rrrr|  rrrr  }
\toprule
  & \multicolumn{4}{c}{$L_1$ Chamfer Distance} & \multicolumn{4}{c}{$L_2$ Chamfer Distance x1000} & \multicolumn{4}{c}{Normal Consistency}\\

  & VP  & CVX & BSP  & Ours & VP & CVX & BSP  & Ours & VP & CVX & BSP  & Ours \\
\midrule

plane           &0.036  &0.023  &0.017  &\textbf{0.011}  &0.997  &0.419  &0.242  &\textbf{0.084}  &0.709  &0.836  &0.791  &\textbf{0.958}\\ 
car             &0.054  &0.023  &0.031  &\textbf{0.019}  &2.157  &\textbf{0.346}  &0.739  &0.363  &0.729  &0.877  &0.675  &\textbf{0.937}\\ 
chair           &0.052  &0.024  &0.027  &\textbf{0.023}  &2.130  &0.769  &\textbf{0.546}  &0.778  &0.694  &0.932  &0.751  &\textbf{0.932}\\ 
lamp           &0.056  &0.023  &0.032  &\textbf{0.020}  &2.747  &\textbf{0.488}  &1.493  &0.823  &0.642  &0.852  &0.692  &\textbf{0.932}\\ 
table           &0.046  &\textbf{0.023}  &0.026  &\textbf{0.023}  &1.690  &\textbf{0.449}  &0.604  &0.510  &0.841  &\textbf{0.938}  &0.803  &\textbf{0.938}\\ 
sofa            &0.057  &\textbf{0.021}  &0.029  &0.023  &2.272  &\textbf{0.283}  &0.675  &0.442  &0.677  &\textbf{0.954}  &0.731  &0.920\\ 
phone          &0.043  &\textbf{0.019}  &0.025  &0.022  &1.382  &\textbf{0.216}  &0.446  &0.346  &0.882  &\textbf{0.966}  &0.820  &0.962\\ 
vessel          &0.045  &0.028  &0.025  &\textbf{0.015}  &1.571  &0.687  &0.558  &\textbf{0.382}  &0.679  &0.802  &0.707  &\textbf{0.935}\\ 
speaker        &0.060  &\textbf{0.027}  &0.041  &0.038  &2.696  &\textbf{0.591}  &1.564  &1.482  &0.730  &\textbf{0.944}  &0.710  &0.869\\ 
cabinet         &0.057  &\textbf{0.026}  &0.036  &0.042  &2.461  &\textbf{0.440}  &1.095  &1.601  &0.730  &\textbf{0.947}  &0.742  &0.846\\ 
display         &0.046  &\textbf{0.021}  &0.025  &0.026  &1.553  &\textbf{0.271}  &0.516  &0.557  &0.859  &\textbf{0.963}  &0.821  &0.937\\ 
bench           &0.043  &0.021  &0.022  &\textbf{0.016}  &1.513  &0.371  &0.389  &\textbf{0.254}  &0.716  &0.870  &0.725  &\textbf{0.936}\\ 
rifle           &0.034  &0.036  &0.016  &\textbf{0.010}  &1.022  &1.514  &0.215  &\textbf{0.070}  &0.712  &0.700  &0.714  &\textbf{0.926}\\ 
\midrule
mean            &0.048  &0.024  &0.027  &\textbf{0.022}  &1.861  &\textbf{0.526}  &0.699  &0.592  &0.738  &0.891  &0.745  &\textbf{0.925}\\ 

\bottomrule
\end{tabular}
}\end{center}

\label{table:compare_baselines}
\vspace{-0.2in}
\end{table*}

\begin{figure}[h]
    \centering
    \includegraphics[width=\textwidth]{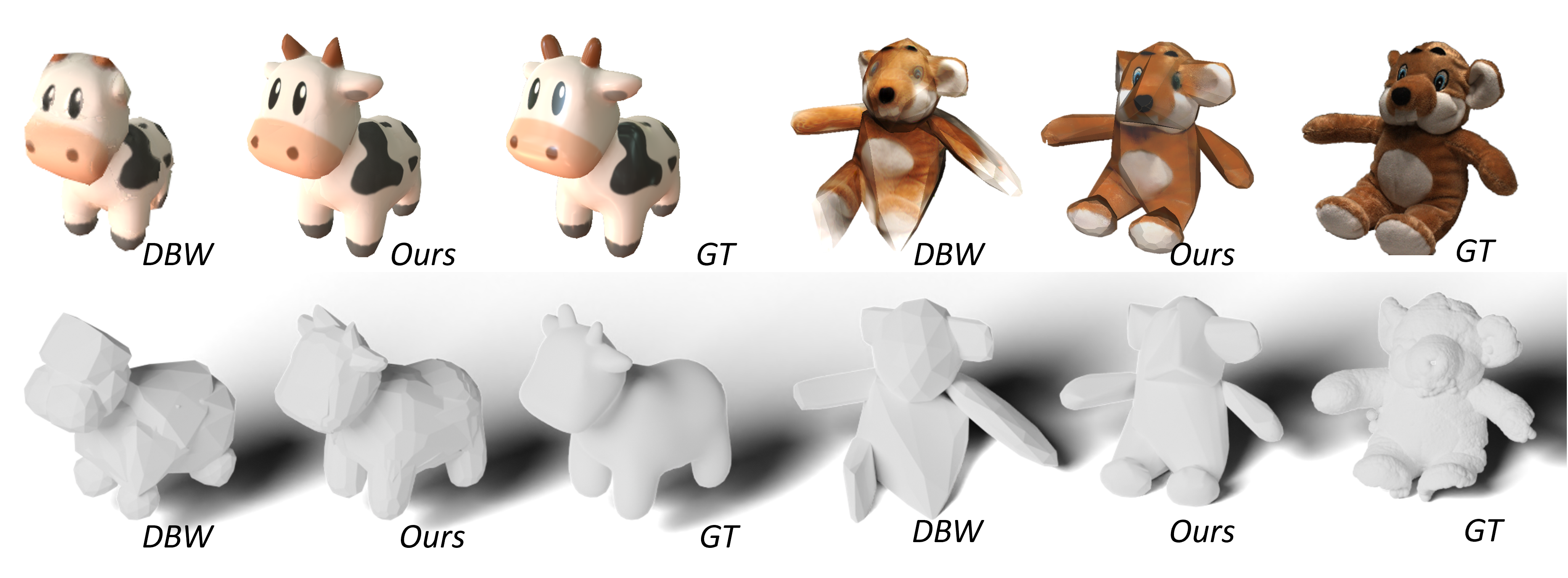}
    \caption{We assess our convex polyhedron-based method against DBW~\cite{monnier2023dbw} that is based on boxes and superquadrics. Visual comparisons demonstrate that our approach yields much better reconstruction results.}
    \label{fig:nvdiffrec}
\end{figure}

\subsection{Multiview Reconstruction}
In this experiment, we extend our model to real world examples, showcasing its effectiveness across various settings and datasets. Given the nature of convex polyhedra, which lack consistent tessellation and triangle topology for a static texture mapping, we adapt the technique from NVDiffRec~\cite{munkberg2022extracting}, applying volumetric textures while solely altering the geometry representation. Our evaluations span both synthetic and real-world captured datasets. We benchmark our approach against models based on other geometric primitives, i.e. DifferentiableBlockWorlds~\cite{monnier2023dbw}. 
We run our method with DTU datasets~\cite{jensen2014large} as well as other synthetic datasets with textures. The quantitative metric of~\cite{monnier2023dbw} on DTU dataset reported in Tab.\ref{tab:nvdiffrec} is from the original paper, and the metric of our method is generated with the same evaluation script from~\cite{monnier2023dbw}. For synthetic datasets (e.g., NeRF Datasets~\cite{mildenhall2020nerf}), we only adopt the superquadric geometry from DBW and use the same pipeline as NVDiffRec~\cite{munkberg2022extracting}, and the metric is computed using $L_1$ Chamfer Distance (multiplied by 10 for better viewing). We show qualitative comparison in Fig.~\ref{fig:nvdiffrec},
which demonstrates better geometry and overall reconstruction quality of our method.

\begin{table}
\caption{Comparison of our method with the method of Differentiable Block Worlds (DBW)~\cite{monnier2023dbw} for multiview reconstruction. We can see that our method outperforms DBW by a large margin.}
\centering
\resizebox{\linewidth}{!}{%
\addtolength{\tabcolsep}{4pt} 
  \begin{tabular}{ l c|ccccc| cc |ccccc}
  \toprule  
  & & \multicolumn{5}{|c|}{DTU} &  \multicolumn{2}{c|}{Mesh} &  \multicolumn{5}{c}{NeRF Synthetic} \\
  & Method & S24 & S40 & S55 & S83 & S105 & Bob&  Spot &  Lego & Chair & Mic & Drums & Hotdog \\
  
  \midrule
 $L_2 \text{ } CD$ & \bf DBW~\cite{monnier2023dbw} &
  \textbf{3.25} & 1.16 & 2.98 & 3.43 & 5.21 &  
  0.71 & 1.03 &  
  0.91 & 1.07 & 1.06 &  1.07 &  1.40  
  \\
  
  $L_2 \text{ } CD$ & \bf Ours &
  3.87 & \textbf{1.01} & \textbf{2.43} & \textbf{2.49} & \textbf{2.62} & 
  \textbf{0.43} & \textbf{0.65} & 
  \textbf{0.44} & \textbf{0.35}&  \textbf{0.25} & \textbf{0.41} & \textbf{0.63} \\
  \bottomrule
  \end{tabular}
}

\label{tab:nvdiffrec}
\end{table}

\begin{wrapfigure}[11]{r}{0.4\textwidth}
  \centering
  \vspace{-20pt}
  \includegraphics[width=0.39\textwidth]{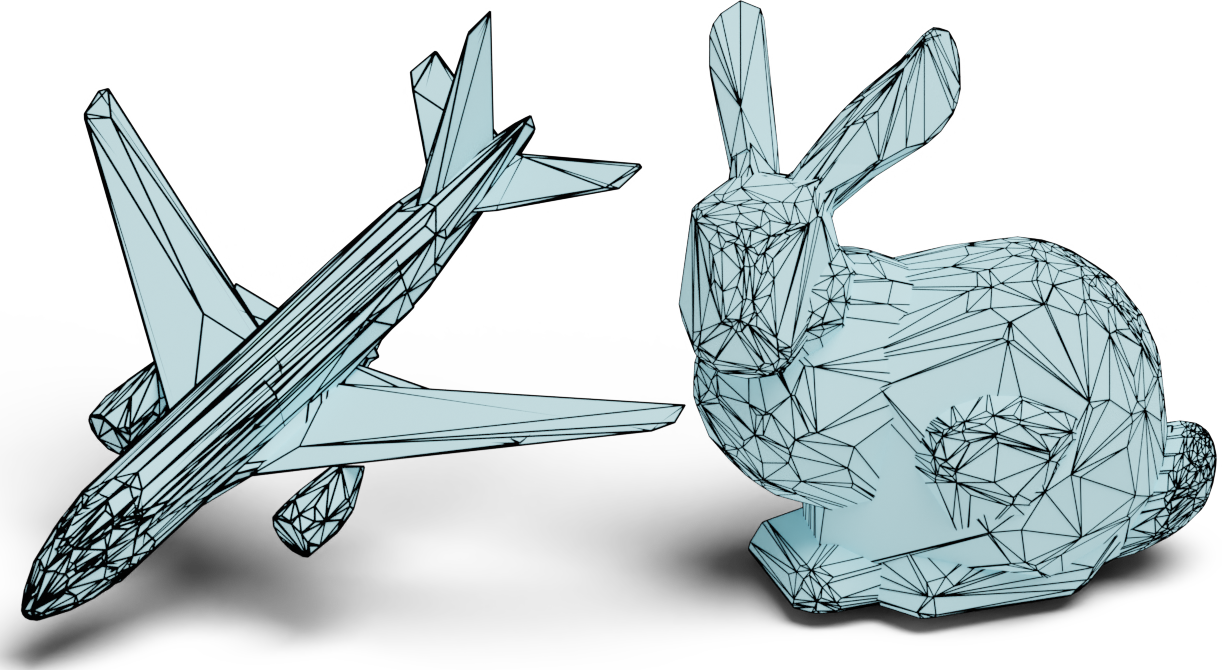}
  \caption{Wireframes of the extracted meshes show the compactness of our outputs.}
  \label{fig:compact} 
\end{wrapfigure}
\subsection{Shape Parsing}
\begin{figure}
    \centering
    \includegraphics[width=0.98\textwidth]{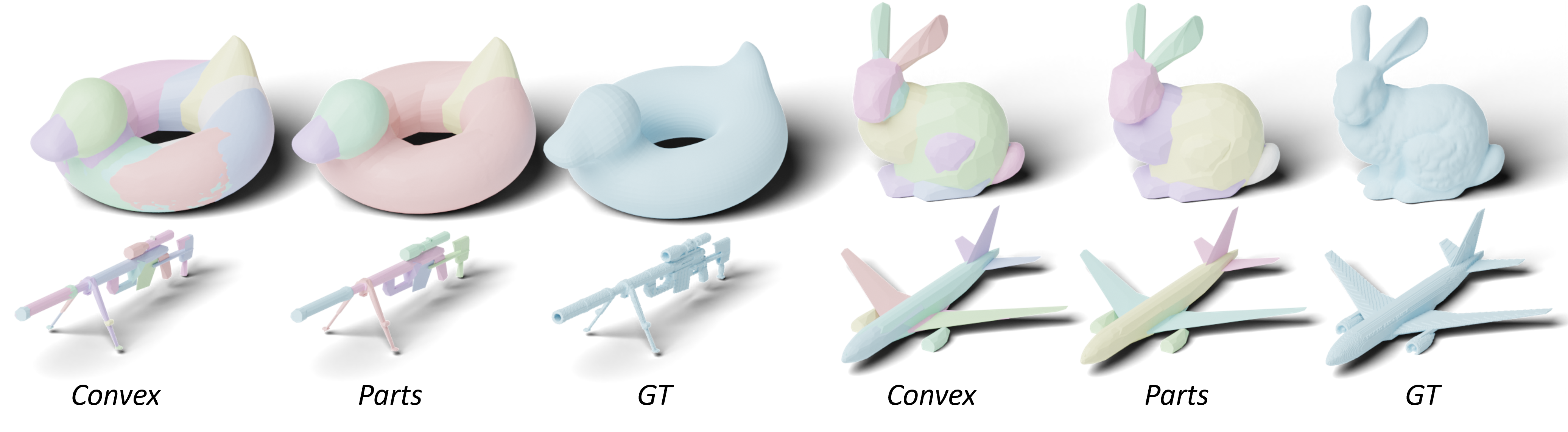}
    \caption{Four objects: Bob, Bunny, Bench, and Airplane. For each object, we show the color-coded individual convex polyhedra on the left and manually grouped object parts in the middle.}
    \label{fig:shape_parsing}
\end{figure}

We have demonstrated the reconstruction ability of our method. Now we delve into a detailed analysis of our reconstruction results, focusing particularly on shape parsing and segmentation capabilities of our method. Our examples span both CAD model types, such as rifles and airplanes, and organic shapes like bunnies and bobs. Fig.~\ref{fig:compact} displays the wire frames of the reconstructed meshes. We can see from the figure that the output mesh is dense in the region with geometric details and sparse in flat regions. We also color-code individual convex polyhedra of the reconstructed shapes, as shown in Fig.~\ref{fig:shape_parsing}. Additionally, we manually group the convexes to illustrate their correspondence with specific parts of the objects, further highlighting our method's adeptness in understanding and segmenting complex shapes.

\subsection{Ablation Study}
An ablation study is conducted to understand the impact of various settings on the performance of our method. Specifically, we explore the effects of convex densification, convex spawning and the number of convexes utilized. Each of these components can play a role in the overall effectiveness and efficiency of shape reconstruction, and understanding their individual and combined contributions is essential for refining the approach. We discuss the effect of each component in the following sections and show numerical results in Tab.\ref{tab:ablation_study}
\begin{figure}
    \centering
    \includegraphics[width=\textwidth]{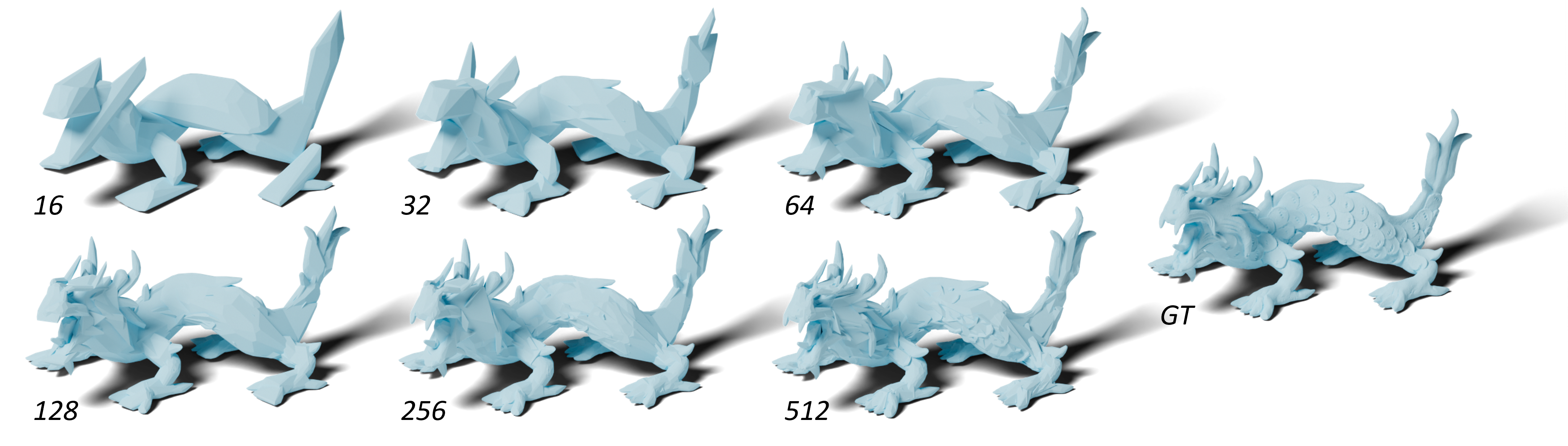}
    \caption{Reconstruction using different numbers of convexes.  A smaller number of available convex polyhedra leads to a higher abstraction level with a more prominent shape parsing structure, while a higher number leads to better reconstruction results, especially on the region with geometry details.}
    \label{fig:ablation_num_convex}
\end{figure}

\begin{wrapfigure}[10]{r}{0.35\textwidth}
  \centering
    \vspace{-20pt}
    \includegraphics[width=0.32\textwidth]{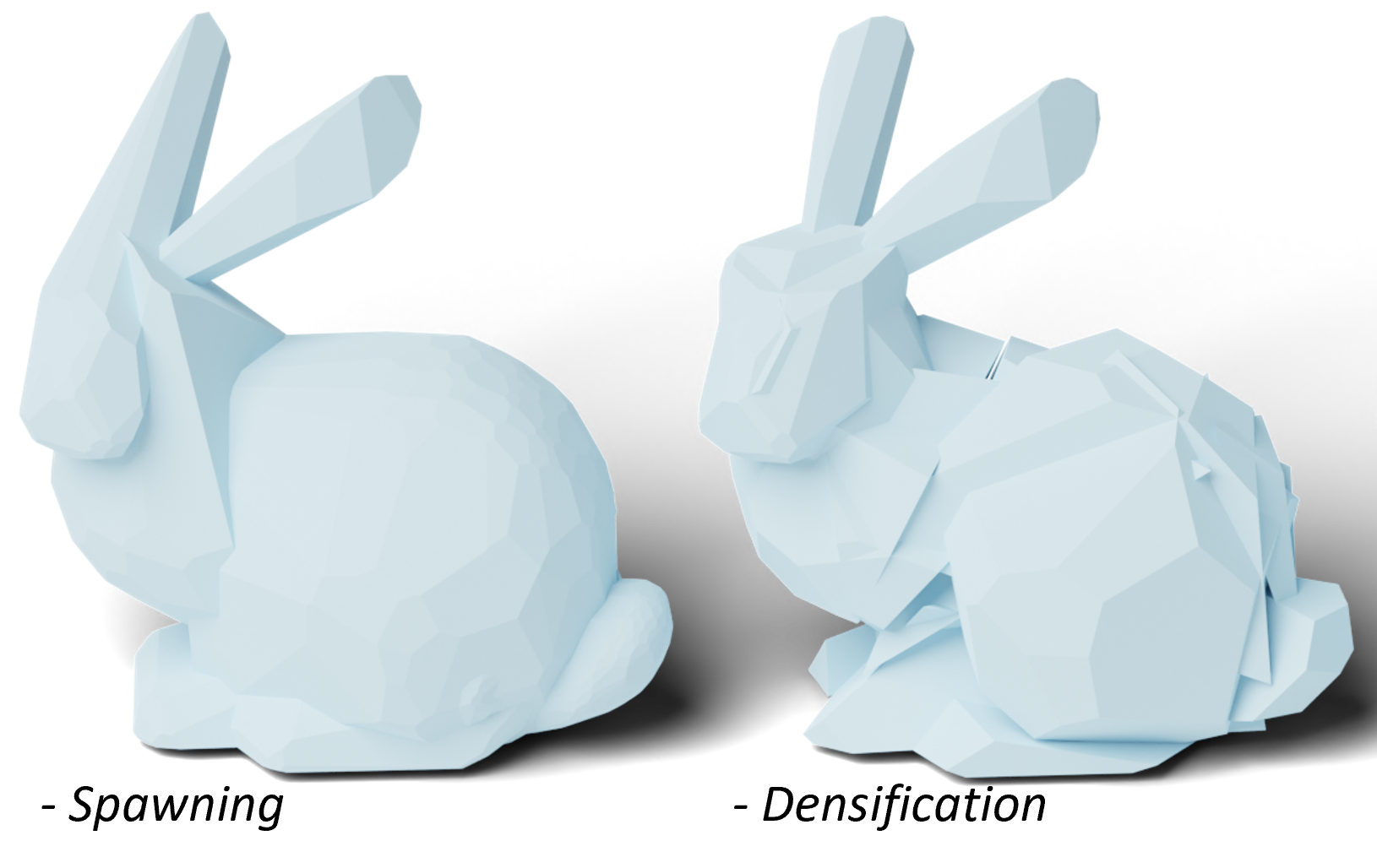}
    \caption{\textbf{Left} without convex spawning, \textbf{Right} without densification process.}
    \label{fig:ablation_initialization_densification_spawning}
\end{wrapfigure}

\noindent\textbf{Number of Convexes}
The number of available convex polyhedra used in the model directly impacts its capacity to reconstruct detailed shapes. We conduct experiments by varying the number of convex polyhedra to visualize the balance between reconstruction accuracy and abstraction levels.  Visual results are given in Fig.~\ref{fig:ablation_num_convex}, where a complex shape (dragon) is optimized with different numbers of convex polyhedra. From the visualization we can see that the higher number gives generally better reconstruction results but with the cost of less abstraction as well as more computation overhead.

\begin{table}
\begin{center}
\resizebox{0.9\linewidth}{!}{
\addtolength{\tabcolsep}{1pt}   
\begin{tabular}{c|cccccc|cc}\toprule
     & 16 Convex   & 32 Convex   & 64 Convex & 128 Convex & 256 Convex & 512 Convex & -Densify$_{16}$ & -Spawn$_{16}$\\ \hline
    CD $L_2$  & 0.32 & 0.14 & 0.08 & 0.05 & 0.05 & 0.05 & 0.46 & 0.54 \\ \bottomrule
    
  \end{tabular}}
\end{center}
\caption{Quantitative results from different optimization configurations.}\label{tab:ablation_study}
\end{table}

\noindent\textbf{Densification}
As discussed in Sec.~\ref{sec:optimization}, we employed a process of adding additional hyperplanes to a convex polyhedron to better capture regions of high curvature or complex detail. We compare models optimized with and without the implementation of convex densification to evaluate its influence on the fidelity of the reconstructed shapes, as shown in Fig.\ref{fig:ablation_initialization_densification_spawning}. We can see from the comparison that the densification process drastically enhances the model's ability to represent curved surfaces, i.e., the heap of the bunny.

\noindent\textbf{Convex Spawning}
In order to maintain the pre-specified number of convexes throughout the optimization process, we re-introduce convex polyhedra during optimization. This process is essential, as poor initial convex polyhedron location can lead to under-reconstruction even when the initial number of convex is large. We refer the readers to Fig.~\ref{fig:ablation_initialization_densification_spawning} for visual comparison.

\section{Conclusion and Limitation}
We have presented a new method for making the optimization of convex polyhedra differentiable w.r.t rendering loss. The key idea is to leverage a non-differentiable duality transform to identify planes intersecting at individual convex polyhedron vertices, which enables the process of solving vertex positions differentiable through the solution of three-plane intersections. The extensive experimentations and detailed ablation studies have demonstrated the effectiveness of our method. Our work will also benefit the research community by offering a new avenue for exploring shape representation with convex polyhedra via differentiable rendering techniques.

\noindent\textbf{Limitation}
Despite its effectiveness, our method has certain limitations. Representing shapes as sets of convex polyhedra results in a loss of detail compared to ordinary meshes and implicit representations. This limitation also restricts current experiments to individual objects rather than entire scenes. The non-static mesh topology prevents predefined parameterization (UV unwrapping) for surface textures, necessitating volumetric textures. Additionally, the densification and purging process is somewhat heuristic. Our method relies on differentiable mesh rendering, inheriting their limitations, such as the lack of gradients for implicit edges (from triangle self-intersections). This can slow down the optimization process and cause it to get stuck in a local minimum. While these limitations do not significantly impact overall reconstruction performance, they highlight areas for further exploration.
\clearpage
\noindent\textbf{Acknowledgements} This work is supported by MOE AcRF Tier 1 Grant of Singapore (RG12/22), and also by the RIE2025 Industry Alignment Fund – Industry Collaboration Projects (IAF-ICP) (Award I2301E0026), administered by A*STAR, as well as supported by Alibaba Group and NTU Singapore. Daxuan Ren was also partially supported by Autodesk Singapore.

\bibliographystyle{splncs04}
\bibliography{main}

\begin{thebibliography}{10}
\providecommand{\url}[1]{\texttt{#1}}
\providecommand{\urlprefix}{URL }
\providecommand{\doi}[1]{https://doi.org/#1}

\bibitem{bangaru2022differentiable}
Bangaru, S.P., Gharbi, M., Luan, F., Li, T.M., Sunkavalli, K., Hasan, M., Bi, S., Xu, Z., Bernstein, G., Durand, F.: Differentiable rendering of neural sdfs through reparameterization. In: SIGGRAPH Asia 2022 Conference Papers. pp.~1--9 (2022)

\bibitem{bangaru2020unbiased}
Bangaru, S.P., Li, T.M., Durand, F.: Unbiased warped-area sampling for differentiable rendering. ACM Transactions on Graphics (TOG)  \textbf{39}(6),  1--18 (2020)

\bibitem{chang2015shapenet}
Chang, A.X., Funkhouser, T., Guibas, L., Hanrahan, P., Huang, Q., Li, Z., Savarese, S., Savva, M., Song, S., Su, H., et~al.: Shapenet: An information-rich 3d model repository. arXiv preprint arXiv:1512.03012  (2015)

\bibitem{chen2020bsp}
Chen, Z., Tagliasacchi, A., Zhang, H.: Bsp-net: Generating compact meshes via binary space partitioning. In: Proceedings of the IEEE/CVF Conference on Computer Vision and Pattern Recognition. pp. 45--54 (2020)

\bibitem{blender}
Community, B.O.: Blender - a 3D modelling and rendering package. Blender Foundation, Stichting Blender Foundation, Amsterdam (2018), \url{http://www.blender.org}

\bibitem{de2000computational}
De~Berg, M.: Computational geometry: algorithms and applications. Springer Science \& Business Media (2000)

\bibitem{deng2020cvxnet}
Deng, B., Genova, K., Yazdani, S., Bouaziz, S., Hinton, G., Tagliasacchi, A.: Cvxnet: Learnable convex decomposition. In: Proceedings of the IEEE/CVF Conference on Computer Vision and Pattern Recognition. pp. 31--44 (2020)

\bibitem{guo20153d}
Guo, K., Zou, D., Chen, X.: 3d mesh labeling via deep convolutional neural networks. ACM Transactions on Graphics (TOG)  \textbf{35}(1),  1--12 (2015)

\bibitem{hao2020dualsdf}
Hao, Z., Averbuch-Elor, H., Snavely, N., Belongie, S.: Dualsdf: Semantic shape manipulation using a two-level representation. In: Proceedings of the IEEE/CVF Conference on Computer Vision and Pattern Recognition. pp. 7631--7641 (2020)

\bibitem{Mitsuba3}
Jakob, W., Speierer, S., Roussel, N., Nimier-David, M., Vicini, D., Zeltner, T., Nicolet, B., Crespo, M., Leroy, V., Zhang, Z.: Mitsuba 3 renderer (2022), https://mitsuba-renderer.org

\bibitem{jensen2014large}
Jensen, R., Dahl, A., Vogiatzis, G., Tola, E., Aan{\ae}s, H.: Large scale multi-view stereopsis evaluation. In: Proceedings of the IEEE conference on computer vision and pattern recognition. pp. 406--413 (2014)

\bibitem{jiang2020sdfdiff}
Jiang, Y., Ji, D., Han, Z., Zwicker, M.: Sdfdiff: Differentiable rendering of signed distance fields for 3d shape optimization. In: Proceedings of the IEEE/CVF conference on computer vision and pattern recognition. pp. 1251--1261 (2020)

\bibitem{kania2020ucsg}
Kania, K., Zieba, M., Kajdanowicz, T.: Ucsg-net--unsupervised discovering of constructive solid geometry tree. arXiv preprint arXiv:2006.09102  (2020)

\bibitem{kerbl20233d}
Kerbl, B., Kopanas, G., Leimk{\"u}hler, T., Drettakis, G.: 3d gaussian splatting for real-time radiance field rendering. ACM Transactions on Graphics  \textbf{42}(4) (2023)

\bibitem{Laine2020diffrast}
Laine, S., Hellsten, J., Karras, T., Seol, Y., Lehtinen, J., Aila, T.: Modular primitives for high-performance differentiable rendering. ACM Transactions on Graphics  \textbf{39}(6) (2020)

\bibitem{li2023secad}
Li, P., Guo, J., Zhang, X., Yan, D.M.: Secad-net: Self-supervised cad reconstruction by learning sketch-extrude operations. In: Proceedings of the IEEE/CVF Conference on Computer Vision and Pattern Recognition. pp. 16816--16826 (2023)

\bibitem{li2018differentiable}
Li, T.M., Aittala, M., Durand, F., Lehtinen, J.: Differentiable monte carlo ray tracing through edge sampling. ACM Transactions on Graphics (TOG)  \textbf{37}(6),  1--11 (2018)

\bibitem{liao2018deep}
Liao, Y., Donne, S., Geiger, A.: Deep marching cubes: Learning explicit surface representations. In: Proceedings of the IEEE Conference on Computer Vision and Pattern Recognition. pp. 2916--2925 (2018)

\bibitem{lien2004approximate}
Lien, J.M., Amato, N.M.: Approximate convex decomposition. In: Proceedings of the twentieth annual symposium on Computational geometry. pp. 457--458 (2004)

\bibitem{lien2007approximate}
Lien, J.M., Amato, N.M.: Approximate convex decomposition of polyhedra. In: Proceedings of the 2007 ACM symposium on Solid and physical modeling. pp. 121--131 (2007)

\bibitem{liu2019softras}
Liu, S., Li, T., Chen, W., Li, H.: Soft rasterizer: A differentiable renderer for image-based 3d reasoning. The IEEE International Conference on Computer Vision (ICCV)  (Oct 2019)

\bibitem{loubet2019reparameterizing}
Loubet, G., Holzschuch, N., Jakob, W.: Reparameterizing discontinuous integrands for differentiable rendering. ACM Transactions on Graphics (TOG)  \textbf{38}(6),  1--14 (2019)

\bibitem{maturana2015voxnet}
Maturana, D., Scherer, S.: Voxnet: A 3d convolutional neural network for real-time object recognition. In: 2015 IEEE/RSJ international conference on intelligent robots and systems (IROS). pp. 922--928. IEEE (2015)

\bibitem{mescheder2019occupancy}
Mescheder, L., Oechsle, M., Niemeyer, M., Nowozin, S., Geiger, A.: Occupancy networks: Learning 3d reconstruction in function space. In: Proceedings of the IEEE/CVF Conference on Computer Vision and Pattern Recognition. pp. 4460--4470 (2019)

\bibitem{mildenhall2020nerf}
Mildenhall, B., Srinivasan, P.P., Tancik, M., Barron, J.T., Ramamoorthi, R., Ng, R.: Nerf: Representing scenes as neural radiance fields for view synthesis. In: European Conference on Computer Vision. pp. 405--421. Springer (2020)

\bibitem{monnier2023dbw}
Monnier, T., Austin, J., Kanazawa, A., Efros, A.A., Aubry, M.: {Differentiable Blocks World: Qualitative 3D Decomposition by Rendering Primitives}. In: {NeurIPS} (2023)

\bibitem{munkberg2022extracting}
Munkberg, J., Hasselgren, J., Shen, T., Gao, J., Chen, W., Evans, A., M{\"u}ller, T., Fidler, S.: Extracting triangular 3d models, materials, and lighting from images. In: Proceedings of the IEEE/CVF Conference on Computer Vision and Pattern Recognition. pp. 8280--8290 (2022)

\bibitem{nichol2022point}
Nichol, A., Jun, H., Dhariwal, P., Mishkin, P., Chen, M.: Point-e: A system for generating 3d point clouds from complex prompts. arXiv preprint arXiv:2212.08751  (2022)

\bibitem{Nicolet2021Large}
Nicolet, B., Jacobson, A., Jakob, W.: Large steps in inverse rendering of geometry. ACM Transactions on Graphics (Proceedings of SIGGRAPH Asia)  \textbf{40}(6) (Dec 2021). \doi{10.1145/3478513.3480501}, \url{https://rgl.epfl.ch/publications/Nicolet2021Large}

\bibitem{nimier2019mitsuba}
Nimier-David, M., Vicini, D., Zeltner, T., Jakob, W.: Mitsuba 2: A retargetable forward and inverse renderer. ACM Transactions on Graphics (TOG)  \textbf{38}(6),  1--17 (2019)

\bibitem{noh2015learning}
Noh, H., Hong, S., Han, B.: Learning deconvolution network for semantic segmentation. In: Proceedings of the IEEE international conference on computer vision. pp. 1520--1528 (2015)

\bibitem{park2019deepsdf}
Park, J.J., Florence, P., Straub, J., Newcombe, R., Lovegrove, S.: Deepsdf: Learning continuous signed distance functions for shape representation. In: Proceedings of the IEEE/CVF Conference on Computer Vision and Pattern Recognition. pp. 165--174 (2019)

\bibitem{paschalidou2019superquadrics}
Paschalidou, D., Ulusoy, A.O., Geiger, A.: Superquadrics revisited: Learning 3d shape parsing beyond cuboids. In: Proceedings of the IEEE/CVF Conference on Computer Vision and Pattern Recognition. pp. 10344--10353 (2019)

\bibitem{preparata2012computational}
Preparata, F.P., Shamos, M.I.: Computational geometry: an introduction. Springer Science \& Business Media (2012)

\bibitem{qi2017pointnet}
Qi, C.R., Su, H., Mo, K., Guibas, L.J.: Pointnet: Deep learning on point sets for 3d classification and segmentation. In: Proceedings of the IEEE conference on computer vision and pattern recognition. pp. 652--660 (2017)

\bibitem{qi2016volumetric}
Qi, C.R., Su, H., Nie{\ss}ner, M., Dai, A., Yan, M., Guibas, L.J.: Volumetric and multi-view cnns for object classification on 3d data. In: Proceedings of the IEEE conference on computer vision and pattern recognition. pp. 5648--5656 (2016)

\bibitem{qi2017pointnet++}
Qi, C.R., Yi, L., Su, H., Guibas, L.J.: Pointnet++: Deep hierarchical feature learning on point sets in a metric space. Advances in neural information processing systems  \textbf{30} (2017)

\bibitem{ravi2020pytorch3d}
Ravi, N., Reizenstein, J., Novotny, D., Gordon, T., Lo, W.Y., Johnson, J., Gkioxari, G.: Accelerating 3d deep learning with pytorch3d. arXiv:2007.08501  (2020)

\bibitem{ren2021csg}
Ren, D., Zheng, J., Cai, J., Li, J., Jiang, H., Cai, Z., Zhang, J., Pan, L., Zhang, M., Zhao, H., et~al.: Csg-stump: A learning friendly csg-like representation for interpretable shape parsing. arXiv preprint arXiv:2108.11305  (2021)

\bibitem{ren2022extrudenet}
Ren, D., Zheng, J., Cai, J., Li, J., Zhang, J.: Extrudenet: Unsupervised inverse sketch-and-extrude for shape parsing. In: European Conference on Computer Vision. pp. 482--498. Springer (2022)

\bibitem{shen2021deep}
Shen, T., Gao, J., Yin, K., Liu, M.Y., Fidler, S.: Deep marching tetrahedra: a hybrid representation for high-resolution 3d shape synthesis. Advances in Neural Information Processing Systems  \textbf{34},  6087--6101 (2021)

\bibitem{shen2023flexible}
Shen, T., Munkberg, J., Hasselgren, J., Yin, K., Wang, Z., Chen, W., Gojcic, Z., Fidler, S., Sharp, N., Gao, J.: Flexible isosurface extraction for gradient-based mesh optimization. ACM Transactions on Graphics (TOG)  \textbf{42}(4),  1--16 (2023)

\bibitem{tulsiani2017learning}
Tulsiani, S., Su, H., Guibas, L.J., Efros, A.A., Malik, J.: Learning shape abstractions by assembling volumetric primitives. In: Proceedings of the IEEE Conference on Computer Vision and Pattern Recognition. pp. 2635--2643 (2017)

\bibitem{Vicini2022sdf}
Vicini, D., Speierer, S., Jakob, W.: Differentiable signed distance function rendering. Transactions on Graphics (Proceedings of SIGGRAPH)  \textbf{41}(4),  125:1--125:18 (Jul 2022). \doi{10.1145/3528223.3530139}

\bibitem{wang2018pixel2mesh}
Wang, N., Zhang, Y., Li, Z., Fu, Y., Liu, W., Jiang, Y.G.: Pixel2mesh: Generating 3d mesh models from single rgb images. In: Proceedings of the European Conference on Computer Vision (ECCV). pp. 52--67 (2018)

\bibitem{wang2021neus}
Wang, P., Liu, L., Liu, Y., Theobalt, C., Komura, T., Wang, W.: Neus: Learning neural implicit surfaces by volume rendering for multi-view reconstruction. arXiv preprint arXiv:2106.10689  (2021)

\bibitem{wang2017cnn}
Wang, P.S., Liu, Y., Guo, Y.X., Sun, C.Y., Tong, X.: O-cnn: Octree-based convolutional neural networks for 3d shape analysis. ACM Transactions On Graphics (TOG)  \textbf{36}(4),  1--11 (2017)

\bibitem{wang2019dynamic}
Wang, Y., Sun, Y., Liu, Z., Sarma, S.E., Bronstein, M.M., Solomon, J.M.: Dynamic graph cnn for learning on point clouds. Acm Transactions On Graphics (tog)  \textbf{38}(5),  1--12 (2019)

\bibitem{wen2019pixel2mesh++}
Wen, C., Zhang, Y., Li, Z., Fu, Y.: Pixel2mesh++: Multi-view 3d mesh generation via deformation. In: Proceedings of the IEEE/CVF International Conference on Computer Vision. pp. 1042--1051 (2019)

\bibitem{wu2022objectsdf}
Wu, Q., Liu, X., Chen, Y., Li, K., Zheng, C., Cai, J., Zheng, J.: Object-compositional neural implicit surfaces. arXiv preprint arXiv:2207.09686  (2022)

\bibitem{xu2022point}
Xu, Q., Xu, Z., Philip, J., Bi, S., Shu, Z., Sunkavalli, K., Neumann, U.: Point-nerf: Point-based neural radiance fields. In: Proceedings of the IEEE/CVF Conference on Computer Vision and Pattern Recognition. pp. 5438--5448 (2022)

\bibitem{yariv2021volume}
Yariv, L., Gu, J., Kasten, Y., Lipman, Y.: Volume rendering of neural implicit surfaces. Advances in Neural Information Processing Systems  \textbf{34},  4805--4815 (2021)

\bibitem{yu2023dualcsg}
Yu, F., Chen, Q., Tanveer, M., Amiri, A.M., Zhang, H.: Dualcsg: Learning dual csg trees for general and compact cad modeling. arXiv preprint arXiv:2301.11497  (2023)

\bibitem{yu2021capri}
Yu, F., Chen, Z., Li, M., Sanghi, A., Shayani, H., Mahdavi-Amiri, A., Zhang, H.: Capri-net: Learning compact cad shapes with adaptive primitive assembly. arXiv preprint arXiv:2104.05652  (2021)

\bibitem{Yu2022MonoSDF}
Yu, Z., Peng, S., Niemeyer, M., Sattler, T., Geiger, A.: Monosdf: Exploring monocular geometric cues for neural implicit surface reconstruction. Advances in Neural Information Processing Systems (NeurIPS)  (2022)

\end{thebibliography}
\end{document}